\documentclass{aa}  

\usepackage[normalem]{ulem}
\usepackage{graphicx}
\usepackage{color}
\usepackage{url}
\usepackage{hyperref}
\usepackage{lipsum}
\usepackage{multirow}
\usepackage{xcolor}

\usepackage{txfonts}
\usepackage{natbib}
\usepackage{amsmath}    
\usepackage{amssymb} 
\usepackage{bm}
\usepackage{mathtools}

\newcommand{\unit}[1]{\mbox{\boldmath $\hat{#1}$}}

\begin{document} 

\title{X-ray polarization from accretion disk winds}

\titlerunning{X-ray polarization from disk winds}

\author{Anagha P. Nitindala\inst{1}
\and
Alexandra Veledina\inst{1,2}
\and
Juri Poutanen\inst{1}
}

\institute{Department of Physics and Astronomy, FI-20014 University of Turku, Finland\\
\email{anaghapradeep.a.nitindala@utu.fi}
\and
Nordita, Stockholm University and KTH Royal Institute of Technology, Hannes Alfv\'ens v\"ag 12, SE-10691 Stockholm, Sweden
}
 
\abstract
{X-ray polarimetry is a fine tool to probe the accretion geometry and physical processes operating in the proximity of  compact objects, black holes and neutron stars.
Recent discoveries made by the Imaging X-ray Polarimetry Explorer put our understanding of the accretion picture in question.
The observed high levels of X-ray polarization in X-ray binaries and active galactic nuclei are challenging to achieve within the conventional scenarios.
In this work we investigate a possibility that a fraction (or even all) of the observed polarized signal arises from scattering in the equatorial accretion disk winds, the slow and extended outflows, which are often detected in these systems via spectroscopic means.
We find that the wind scattering can reproduce the levels of polarization observed in these sources.
}

\maketitle


\section{Introduction}

Accretion onto compact objects, such as neutron stars (NSs) and black holes (BHs), fuels some of the brightest X-ray sources: X-ray binaries (XRBs) and active galactic nuclei (AGN). 
In XRBs, mass is accreted onto the compact object from a close companion. 
They are found in different spectral states: hard and soft states in XRBs hosting BHs, and banana and island states in NS XRBs \citep{ZG04,RemillardMcClintock2006,Done07}, and these states are believed to be associated with different accretion geometry.
The spectral changes are linked to the changes in the dominant radiative mechanism responsible for the broadband spectral production.

The hard-state spectra of BH XRBs are dominated by the Comptonization of soft seed photons by hot electrons in an optically thin inner flow or corona \citep{ST80,PS96,PKR97,Esin97}; the soft state is dominated by a blackbody-like emission produced by a geometrically thin, optically thick accretion disk \citep{SS73,NT73,PT1974}.
Hard- and soft-intermediate states are also identified at the transitions between these two major states \citep{Homan2005}.
Similar to XRBs, X-ray spectra of accreting supermassive BHs in AGN (in particular, Seyfert galaxies considered here) are dominated by a power-law-like component (due to Compton upscattering of soft photons), with a high energy cut-off and the multi-temperature disk blackbody peaking in ultraviolet wavelengths \citep{NandraPounds1994,Pounds1995,ZPJ00}.
The current understanding of the accretion geometry and its underlying physical mechanisms in XRBs and AGN is still incomplete.
The location of the hot Comptonizing medium \citep{Poutanen2018,Bambi2021} and the structure and stability of the optically thick accretion disk \citep{DexterQuataert2012,Jiang2012} remain under debate. 

X-ray polarimetry is a novel tool that can provide independent estimates of the accretion geometry, parameters of compact object, and radiation mechanisms.
The polarization angle (PA) is related to the global axis in the system, such as accretion disk, jet or BH/NS spin: intrinsic polarization is expected to be either aligned or orthogonal to this axis.
Effects of general and special relativity may lead to deviations of PA from this axis \citep[or from the direction orthogonal to that;][]{ConnorsStark1977,StarkConnors1977,Pineault1977pol_schw,PineaultRoeder1977kerr_num,Dovciak2008,SchnittmanKrolik2009,SchnittmanKrolik2010,Loktev2022,Loktev2024}.
The polarization degree (PD) generally depends on the mechanism producing the broadband spectra. 
Pure electron-scattering optically thick disk atmospheres, that can be relevant to soft-state spectra, give maximal PD$=11.7$\% for the edge-on observers \citep{Cha60,Sob63}. 
For the case of Comptonization, the maximal PD$\sim$20\% can be achieved in the slab geometry \citep[][somewhat higher PD of the scattered component can be achieved for the case of Thomson scattering in a slab, \citealt{ST85}]{PS96}. 
PD tends to zero as the corona geometry becomes spherical, but a non-zero net polarization is expected in this case once reflection from the disk is taken into account \citep{Matt1993,PNS96,Dovciak2004}.
The PD is generally expected to depend on the observer inclination: its maximal value is achieved when the source is viewed edge-on; PD decreases to zero for the face-on systems.

Given uncertainty of the accretion geometry, value of BH spin and potential variability of the central source, leading to depolarization, as well as the typically low inclinations of many sources in the sample, general expectations for PD in the hard-state systems were low, typically falling below 2\% \citep{KrawczynskiBeheshtipour2022,ZhangDovciak2022}.
Nevertheless the X-ray PA, when compared to the jet direction, could discriminate between different alternative geometries.
Soft-state data, in turn, carried promises to probe the BH spin via the rate of depolarization due to PA rotation across the energy band \citep{StarkConnors1977,ConnorsPiranStark1980,Dovciak2008,Loktev2022,Loktev2024}.

%
%

The first X-ray polarimetric observations of XRBs by the Imaging X-ray Polarimetry Explorer \citep[IXPE;][]{Weisskopf2022} have led to puzzling results \citep[see][for reviews]{Dovciak2024,Poutanen2024Galax,Ursini2024}. 
Low-inclination BH binaries in the hard(/hard-intermediate)-state were found highly polarized \citep[PD$\approx$4\% in Cyg~X-1 and Swift~J1727.8--1613;][]{Krawczynski22,Veledina2023,Ingram2024,Podgorny2024}, in line with the constraints on the AGN \citep{Gianolli2023,Ingram2023}; in some sources PD reached exceptionally high values $\sim10-20$\% \citep{Ursini2023,Veledina2024}. 
Soft(/soft-intermediate)-state sources showed no signs of PA rotation \citep{Svoboda2024lmcx-3,Marra2024,Ratheesh2024,Steiner2024,Veledina2024soft} and often a high PD, exceeding standard expectations for the known inclinations, was found.
In a number of NS XRBs, misalignment of PA from the jet axis or its rotation was detected \citep{Doroshenko2023,Rankin2024,LaMonaca2024,Bobrikova2024,Bobrikova2024new}. 

These observational properties are challenging to address within the conventional framework describing the emission of polarized radiation from such sources.
Several adjustments to the original models and geometries were considered to better align their predictions with the data.
Misalignment between the orbital axis and the spin of the compact object was invoked to enhance the PD and to alter the polarization axis \citep{Krawczynski22,Bobrikova2024,Rankin2024}.
A higher X-ray PD could be obtained for the cases of an outflowing material, due to the aberration effect \citep{Poutanen2023,Ratheesh2024,DexterBegelman2024,Sridhar2024}. 
In many cases, however, explaining the observed properties requires stretching or fine-tuning the parameters.  
It also remains unclear why the soft-state sources do not show any signs of PA rotation.

An alternative approach to the problem is to assume that polarization originates (partially, entirely or occasionally) far from the central source, remaining unaffected by the influences of strong gravity and fast motions of matter.
A natural site for polarization production is scattering of emission from the central source by accretion disk winds.

Observational signatures of accretion disk winds and outflows such as P-Cyg line profiles, blue-shifted lines, absorption troughs, and broad-emission-line wings have been ubiquitously found in XRB spectra \citep{NeilsenLee2009,Ponti2012,Ponti2016,Munoz-Darias2016,Munoz-Darias2020,DiazTrigo2016,MataSanchez2018}.
These signatures have been detected throughout the entire outbursts of XRBs in various wavelengths from near-infrared to X-rays \citep{Sanchez-Sierras2020,Munoz-DariasPonti2022,CastroSegura2022,Parra2024},
suggesting that outflows might be common and always present in X-ray binaries. 
Furthermore, non-zero optical linear polarization of XRBs is known to accompany the wind detections and may originate from the scattering in an optically thin wind \citep{Kosenkov2017,Veledina2019,Kosenkov2020,Kravtsov2023atel}.
This leads to a suggestion that accretion disk winds could play a role in contributing to X-ray polarization in XRBs.
Previous study nevertheless gave negative results \citep{Tomaru2024}.

In the case of broad absorption-line quasars and radio-quiet AGN, signatures of winds have been found through absorption troughs and blue-shifted lines in UV and X-ray wavelengths \citep{2009ApJ...692..758G,2010A&A...521A..57T}. Narrow blue-shifted absorption lines indicating outflows have been seen across various ionization states \citep{2003ARA&A..41..117C,2010SSRv..157..265C}. Equatorial accretion disk winds have also been assumed to explain the wings of broad absorption-line quasars \citep{1992ApJ...385..460E,1995ApJ...451..498M}. Winds in AGN have been a popular model to unify the different types of AGN -- low luminosity AGN, Seyferts, and broad absorption-line quasars \citep{2000ApJ...545...63E,2019A&A...630A..94G}. Thus, it is likely that accretion disk winds is a common feature in AGN as well, similar to XRBs.

In this paper, we present a broad study of the effects of accretion disk winds on the X-ray polarimetric properties of XRBs and AGN. 
We consider various properties of the accretion disk wind, such as opening angle and optical depth, and angular properties and nature of the illuminating central source. 
The paper is organized as follows.
The model and the method of solution are described in Sect.~\ref{sec:setup}. 
We present the results in Sect.~\ref{sec:results}. 
We then discuss our findings and compare the results to the data obtained with IXPE on NS and BH XRBs and Seyfert 1 galaxies in Sect.~\ref{sec:discuss}. 
We summarize our findings in Sect.~\ref{sec:summary}.

\section{Model setup and method}
\label{sec:setup}

\begin{figure}
\centering
\includegraphics[clip=true, trim={20 25 0 0 },width=\linewidth]{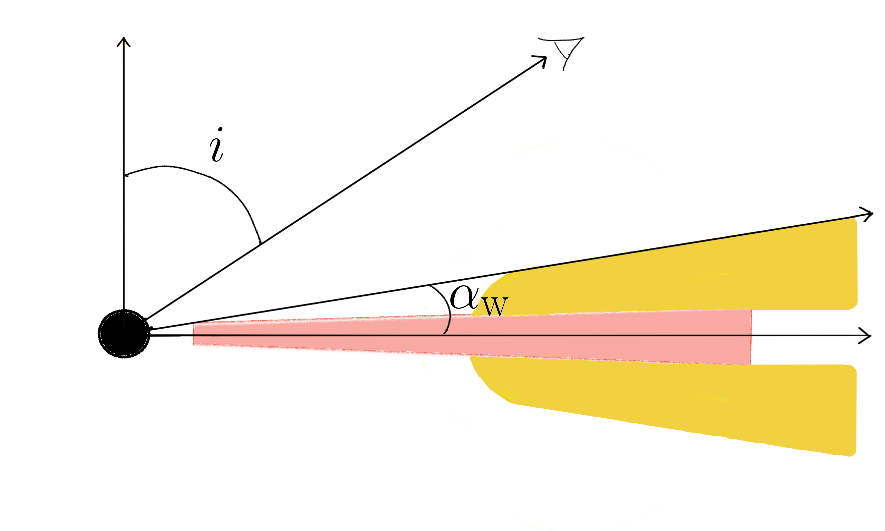}
\caption{Geometry of the system. 
The central point-like illuminating source is described by the black circle while the winds are shown in yellow and the disk in red. 
The observer is at an inclination angle $i$. 
The opening angle of the wind is given by $\alpha_{\rm w}$. 
}
\label{fig:geometry}
\end{figure}

We consider a problem of scattering of the X-rays produced by the central source in an extended accretion disk wind (see Fig.~\ref{fig:geometry}). 
For our application, we can  assume that the central X-ray source is point-like.
This is a valid assumption because the size of the X-ray source of a few Schwarzschild radii is much smaller than the characteristic size of the wind. 
The radiation from the central source is described by the luminosity per unit solid angle and its energy dependence, and, in a general case, it depends on the angle relative to the disk axis and on the azimuthal angle, and it also can be polarized.
The polarization vector is not necessary aligned with the projection of the symmetry axis on the sky or is perpendicular to that.   



\subsection{Radiative transfer equation in a slab}

The radiation field is fully described by the Stokes vector \citep{1851TCaPS...9..399S} $\vec{I}=(I,Q,U,V)^{\rm T}$. 
We introduce the spherical coordinate system with the z-axis along the disk axis $\unit{n}=(0,0,1)$.   
We choose the external polarization basis defined by the vector $\unit{n}$ and the unit vector along the photon momentum $\unit{k}=(\sqrt{1-\mu^2}\cos\phi,\sqrt{1-\mu^2}\sin\phi,\mu)$, with  $\mu$  being the cosine of the polar angle  and $\phi$ is the azimuthal angle.  
The basis for the scattered photon  is given by 
\begin{eqnarray} \label{eq:polbas_k}
\unit{e}^{\rm ex}_1 (\unit{k}) & =&  \frac{\unit{n}-\mu\  \unit{k}}{\sqrt{1-\mu^2}} =  
\left(-\mu\cos\phi, - \mu\sin\phi, \sqrt{1-\mu^2}\right),\\
\unit{e}^{\rm ex}_2 (\unit{k}) & = & \frac{\unit{k} \times \unit{n}}{\sqrt{1-\mu^2}} = (\sin\phi,-\cos\phi,0).
\end{eqnarray} 
Similar expressions can be written  for the external basis for the incoming photon before scattering $\unit{k}'=(\sqrt{1-\mu'^2}\cos\phi',\sqrt{1-\mu'^2}\sin\phi',\mu')$ with changing $\phi$ by $\phi'$ and $\mu$ by $\mu'$. 

Scattering of radiation is best described using the internal bases formed by the two photon momenta. 
The internal basis for the scattered photon is
\begin{equation} \label{eq:polbas_scat}
\unit{e}^{\rm in}_1 (\unit{k}) = \frac{\unit{k}'-\cos\theta\  \unit{k}}{\sin\theta},\qquad
\unit{e}^{\rm in}_2 (\unit{k})  = \frac{\unit{k} \times \unit{k}'}{\sin\theta} , 
\end{equation}
where $\cos\theta= \mu\mu' + \sqrt{1-\mu^2}\sqrt{1-\mu'^2 }  \cos(\phi-\phi')$ is the cosine of the scattering angle. 
Similarly, the internal basis for the incoming photon can be obtained by replacing $\unit{k} \leftrightarrow \unit{k}'$.

Transformation of the Stokes vector between the external and internal bases is described by the rotation matrix 
\begin{equation}
\label{eq:rte_rotation}
\tens{L} (\chi) =  \begin{bmatrix}   
1 & 0 & 0 & 0 \\ 
0 & \cos2\chi & \sin2\chi & 0 \\ 
0 & -\sin2\chi & \cos2\chi & 0 \\ 
0 & 0 & 0 & 1 \\ 
\end{bmatrix}  , 
\end{equation}
where for scattered photon $\unit{k}$ we have 
\begin{eqnarray}
\label{eq:coschi}
\cos\chi  & =&  \unit{e}^{\rm ex}_1 \cdot \unit{e}^{\rm in}_1 = \unit{e}^{\rm ex}_2 \cdot \unit{e}^{\rm in}_2 =  
\frac{\mu' - \mu \cos\theta }{\sin\theta \sqrt{1-\mu^2}}, \\ 
\sin\chi  & =&  \unit{e}^{\rm ex}_2 \cdot \unit{e}^{\rm in}_1 = - \unit{e}^{\rm ex}_1 \cdot \unit{e}^{\rm in}_2 = 
- \frac{\sqrt{1-\mu'^2}\sin(\phi'-\phi)}{\sin\theta } . 
\end{eqnarray} 
Similar expressions can be written for an incoming photon $\unit{k}'$. 

A rather general form of the radiative transfer equation (RTE) describing Thomson scattering (and true absorption) in the plane-parallel atmosphere that allows us to consider arbitrary polarized source of radiation as well as non-axisymmetric scattering medium can be written as  \citep{Cha60}: 
\begin{equation}
\label{eq:rte_general}
\mu \frac{d\vec{I} (\tau, \mu, \phi)}{d\tau} = 
- \frac{1}{\lambda}\vec{I} (\tau, \mu, \phi) + 
\vec{S}(\tau, \mu, \phi),
\end{equation}
where $d\tau=\sigma_{\rm T}n_{\rm e}dz$ is the Thomson scattering optical depth,  $\lambda$ is the albedo for single scattering (i.e. the ratio of the scattering to the total opacity).
The source Stokes vector $\vec{S}$ can be expressed through the integral over the solid angle of the Stokes vector of the incoming radiation and the phase matrix: 
\begin{equation}\label{eqn:source}
\vec{S}(\tau, \mu, \phi) =   \int\limits_{-1}^{1}  d\mu ' \int\limits_0^{2\pi}d\phi' \ \tens{P}(\mu, \phi; \mu', \phi')\vec{I} (\tau, \mu', \phi'), 
\end{equation} 
where the phase matrix is given by the product of two rotation matrices and the scattering matrix:  
\begin{equation}
\tens{P}(\mu, \phi; \mu ', \phi ')  = \tens{L}(-\chi) \tens{R}(\theta) \tens{L}(\chi_1) .
\end{equation} 
The Thomson scattering matrix is 
\begin{equation}
\label{eq:rte_thomson}
\tens{R} (\theta) = \frac{3}{16\pi} \begin{bmatrix}
1+\cos^2\theta & -\sin^2\theta & 0 & 0 \\ 
-\sin^2\theta & 1+\cos^2\theta & 0 & 0 \\ 
0 & 0 & 2\cos\theta & 0 \\ 
0 & 0 & 0 & 2\cos\theta \\ 
\end{bmatrix}    
\end{equation} 
and the elements of the phase matrix are  
\begin{eqnarray}
P_{11}& = & R_{11} , \nonumber \\ 
P_{12}& = & R_{12}\cos2\chi_1 , \nonumber \\ 
P_{13}& = &  R_{12}\sin2\chi_1 , \nonumber \\ 
P_{21}& = & R_{12}\cos2\chi , \nonumber  \\
P_{31}& = & R_{12}\sin2\chi ,\\  
P_{22}& = & R_+ \cos2(\chi-\chi_1) + R_- \cos2(\chi+\chi_1) , \nonumber \\  
P_{23}& = & - R_+ \sin2(\chi-\chi_1) + R_- \sin 2(\chi+\chi_1) , \nonumber \\ 
P_{32}& = & R_+ \sin2(\chi-\chi_1) + R_- \sin2(\chi+\chi_1) , \nonumber \\ 
P_{33}& = & R_+ \cos2(\chi-\chi_1) - R_- \cos2(\chi+\chi_1) , \nonumber \\  P_{44}& = & R_{44} , \nonumber 
\end{eqnarray} 
where $R_{\pm}=\frac{3}{32\pi}(1\pm\cos\theta)^2$ and other matrix elements are zeros. 
Cosines and sines of the difference and sum of $\chi$ and $\chi_1$ can be computed using relations 
\begin{eqnarray} 
\pm \cos(\chi\pm \chi_1) (1\mp\cos\theta) & =& \sqrt{1-\mu^2}\sqrt{1-\mu'^2} \nonumber \\ 
&+&  (\mu\mu'\mp1) \cos(\phi'-\phi), \\ 
\pm \sin(\chi\pm \chi_1) (1\mp\cos\theta) & =& (\mu'\mp \mu) \sin(\phi'-\phi). 
\end{eqnarray}

Let us consider a slab of finite vertical (scattering) optical depth $\tau_0$ with the boundary condition at the bottom $\vec{I} (\tau=0, \mu, \phi)=\vec{B}(\mu, \phi)$.    
For small $\tau_0$, we can solve the RTE in the single-scattering approximation by representing the solution as a sum of unscattered radiation and radiation that is scattered once. 
The unscattered radiation is: 
\begin{equation}
 \vec{I}_0 (\tau, \mu, \phi) =  \vec{B}(\mu, \phi) \exp(-\tau/\mu\lambda) .
\end{equation}
The single-scattered radiation is then given by the Stokes vector 
 \begin{equation}
  \vec{I}_1 (\tau, \mu, \phi) =  \int\limits_0^{\tau} \frac{d\tau'}{\mu} 
  \vec{S}_1 (\tau',\mu,\phi) \exp[-(\tau-\tau')/\lambda\mu], 
 \end{equation}
where $\vec{S}_1$ is given by Eq.~\eqref{eqn:source} with $\vec{I}_0$ instead of $\vec{I}$. 
In the case of small optical depth (when single-scattering approximation can be used), $\vec{I}_1$ can be approximated 
 \begin{equation}\label{eq:I1_app}
  \vec{I}_1 (\mu, \phi) \approx   
  \frac{\tau}{\mu}  \int\limits_{-1}^{1}  d\mu '   \int\limits_0^{2\pi} d\phi' \ \ \tens{P}(\mu, \phi; \mu', \phi')\vec{B} (\mu', \phi') .  
 \end{equation}
The same expression can be rewritten in terms of the luminosity per unit solid angle emitted in a given direction $\vec{L}\propto \mu\vec{I}$ as 
 \begin{equation}\label{eq:lum1_app}
\vec{L}_1 (\mu, \phi)   \approx   
 \int\limits_{-1}^{1} d\mu '   \int\limits_0^{2\pi}  d\phi' \ \  \tens{P}(\mu, \phi; \mu', \phi')  \vec{L}_{\star} (\mu', \phi') \frac{\tau}{\mu'} ,  
 \end{equation} 
 where the incident luminosity at  the boundary is $\vec{L}_{\star}(\mu,\phi)\propto \mu \vec{B}(\mu,\phi)$ and the unscattered escaping luminosity 
\begin{equation}\label{eq:lum0}
 \vec{L}_0(\mu, \phi)=\vec{L}_{\star}(\mu, \phi) \exp(-\tau/\mu\lambda). 
  \end{equation}
The total escaping luminosity (Stokes vector) is   
 \begin{equation}
  \vec{L} (\tau, \mu, \phi) =    \vec{L}_0 (\tau, \mu, \phi)+   \vec{L}_1 (\tau, \mu, \phi). 
 \end{equation}

\begin{figure*}
\centering
\includegraphics[width=0.8\linewidth]{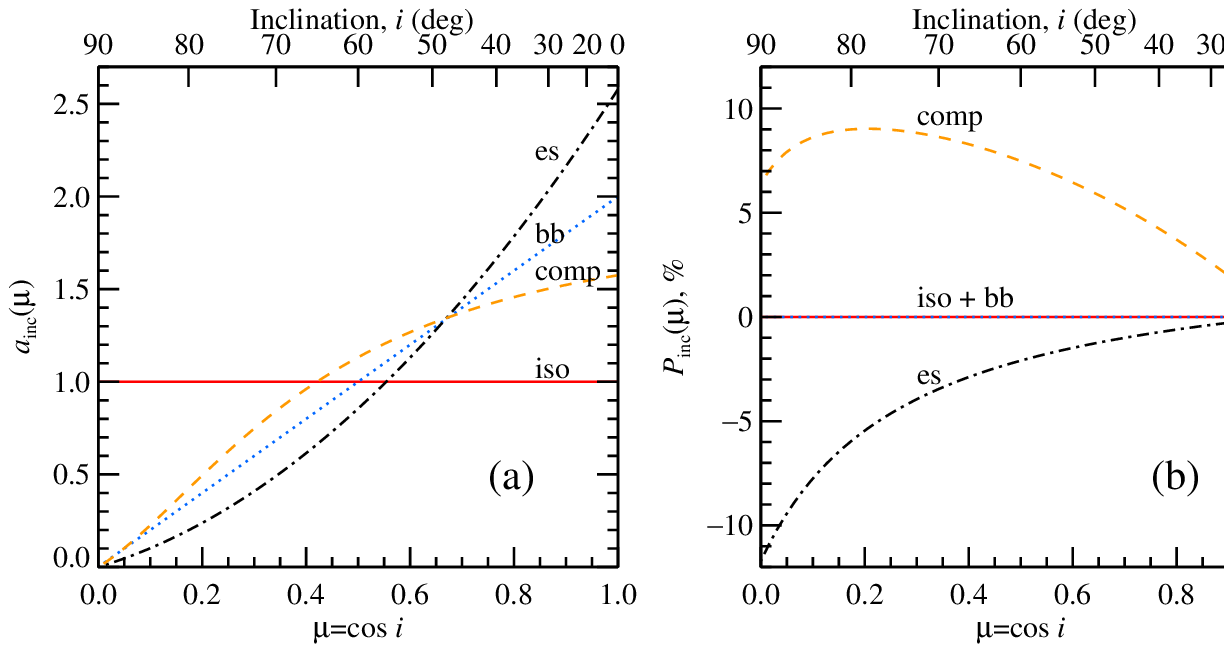}
\caption{Angular distribution (a) and polarization (b) of the incident radiation. 
The solid red, dotted blue, dashed orange, and dot--dashed black lines correspond to the isotropic (iso), blackbody (bb), Comptonization in a slab (comp), and electron-scattering-dominated disk (es) cases, respectively. }
\label{fig:lum_pol}
\end{figure*}

\subsection{Scattering in a wind} 

Scattering geometry of the wind is  obviously not a slab. 
However, the expressions for the escaping luminosity derived above hold with a simple substitution of $\tau/\mu'$ in Eq.~\eqref{eq:lum1_app} by the probability for a photon emitted by the central source at angle $\mu'$ to be scattered within the wind, which is just the scattering optical depth through the wind in a given direction.  

In the following we consider the central illuminating source and the scattering medium to be independent of the azimuthal angle. 
For the radiation source this means that polarization is either parallel to the symmetry axis or is perpendicular to that, thus the Stokes $U = 0$. 
Hence, even though the phase matrix in Eq.~\eqref{eqn:source} depends on the azimuthal angles of the incoming ($\phi'$) and scattered ($\phi$) radiation, its integration over the azimuthal angle from $0$ to $2\pi$ yields Stokes $U = 0$ for the scattered radiation too. 
Moreover, below we will only consider linearly polarized source, so that the Stokes $V = 0$ as well. 
We also assume pure scattering, i.e. $\lambda=1$. 

The Stokes vector thus contains only two components ($I,Q$). 
The unscattered radiation is given by equation similar to Eq.~\eqref{eq:lum0}: 
\begin{equation}\label{eq:lum0_mod}
\vec{L}_0(\mu)=\vec{L}_{\star} (\mu)\ \exp[-\tau(\mu)] 
\end{equation}
and the single-scattered radiation is 
\begin{equation}\label{eqn:source_lum}
\vec{L}_1 (\mu)  =   
\int\limits_{0}^{1}  d\mu'  \ \  \tens{P}(\mu;\mu')  \vec{L}_{\star} (\mu')\  \{1-\exp[-\tau(\mu')]\} ,  
\end{equation} 
where the $2\times2$  phase matrix \citep{Cha60}  
\begin{equation} \label{eq:phasematrix}
\tens{P}(\mu; \mu') = \frac{3}{16}
\begin{bmatrix}
3 - {\mu'}^2 - \mu^2 + 3{\mu'}^2\mu^2 & (1 - {\mu'}^2)(1 - 3\mu^2) \\
(1 - \mu^2)(1 - 3{\mu'}^2)           & 3(1 - {\mu'}^2)(1 - \mu^2) 
\end{bmatrix}
\end{equation}
now depends only on $\mu$ and $\mu'$. 
For small optical depth through the wind, the scattering probability is just $1-\exp[-\tau(\mu')]\approx\tau(\mu')$. 
We ignore here multiple scatterings. 
We note also that the integration limits in Eq.~\eqref{eqn:source_lum} imply that only radiation emitted to the upper hemisphere and scattered in the wind reaches the observer. 
In reality, for the wind extending to distances exceeding disk outer radius, there could be a contribution of photons emitted to the lower hemisphere. 
This additional source would increase the strength of the scattered, polarized signal. 


\subsection{Source of illuminating  radiation} 

An important physical aspect of the system is the property of  the central illuminating  source luminosity, which we represent as 
\begin{equation}\label{eqn:lstar}
\vec{L}_{\star}(\mu) = \frac{L_{\star}}{4\pi} 
             a_{\rm inc}(\mu)
        \begin{bmatrix}
        1 \\
    P_{\rm inc}(\mu)
        \end{bmatrix} , \quad \mbox{for}\ \mu>0, 
\end{equation}
where $L_{\star}$ (without arguments) is the total source luminosity, the angular distribution normalized to unity is $\int_0^1 a_{\rm inc}(\mu)d\mu=1$, and $P_{\rm inc}(\mu)$ is polarization of incident radiation. 
We consider four cases describing different possible distributions. 
First, we consider the simplest case of an isotropic unpolarized source: 
\begin{equation}
a_{\rm iso}(\mu)=1 , \quad 
P_{\rm iso}(\mu)=0 . 
\end{equation}

The second case considered here corresponds to the blackbody radiation from a flat disk.
While the intensity does not depend on viewing angle, the angular distribution of incident luminosity is proportional to the cosine of the inclination: 
\begin{equation}
a_{\rm bb}(\mu)=2\mu, \quad 
P_{\rm bb}(\mu)=0 . 
\end{equation}
These two cases can be considered as the first approximation to the incident X-ray emission of accreting BHs in the hard and soft state, respectively. 
Similarly, the NS spreading layer and the boundary layer/accretion disk can also be approximated by the isotropic spherical and flat sources, respectively.
 
However, the sources of incident X-ray emission in binaries do not need to be unpolarized or isotropic. 
The third case we consider is a model that maybe applicable to the hard state of accreting BH X-ray binaries where X-ray radiation is believed to be produced by thermal Comptonization. 
In this case, we take the angular distribution of radiation and polarization degree at 4~keV (in the middle of IXPE energy range) produced by Comptonization in a static slab (model B in \citealt{Poutanen2023}) of Thomson optical depth $\tau_{\rm T}=1$ and electron temperature of $kT_{\rm e}=100$~keV. 
The angular dependencies in this case can be approximated 
with  
\begin{equation} 
\begin{aligned} 
 a_{\rm comp}(\mu) & =  1.73\ \mu \  \frac{1+5.3\mu-0.2\mu^2}{1+1.3\mu+4.4\mu^2} ,  \\ 
 P_{\rm comp}(\mu) & =    0.064\  (1-\mu)\   \frac{1+16.3\mu+6.2\mu^2}{1+8.2\mu-2.1\mu^2} .  
\end{aligned}
\end{equation}
The positive $P$ means that polarization vector lies in the meridional plane formed by the normal to the slab and the photon momentum. 

The last considered model is the pure electron-scattering dominated atmosphere that may describe properties of the accretion disk in the soft state of BH X-ray binaries. 
The angular distribution and polarization can then be approximated by simple formulae \citep{Cha60,Sob63,ViironenPoutanen2004,SPW20}: 
\begin{equation} 
\begin{aligned} 
 a_{\rm es}(\mu)  & =  2\mu\  \frac{1+2.06\mu}{1+(2/3) 2.06} , \\
P_{\rm es}(\mu) & =  -0.117 \ \left(\frac{1-\mu}{1+3.582\mu} \right ) ,
\end{aligned}
\end{equation}
with the minus sign implying polarization being perpendicular to the disk normal. 
The four cases of the angular distributions of $a_{\rm inc}(\mu)$ and $P_{\rm inc}(\mu)$ are shown in Fig.~\ref{fig:lum_pol}.

\begin{figure}
\centering
\includegraphics[width=0.85\linewidth]{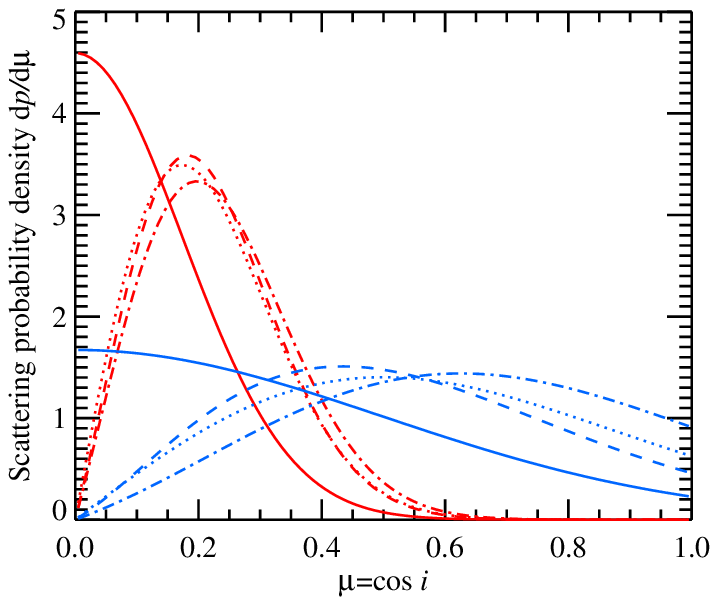}
\caption{Probability density for photons to be scattered in a wind for different emission models. 
The solid, dotted, dashed, and dot--dashed lines correspond to the cases isotropic, blackbody, Comptonization, and electron-scattering-dominated disk, respectively.  
The red  and blue lines correspond to the cases of the wind opening angle $\alpha_{\rm w}=10\degr$ and 30\degr, respectively.}
\label{fig:dpdmu}
\end{figure}

\begin{figure}
\centering
\includegraphics[width=0.9\linewidth]{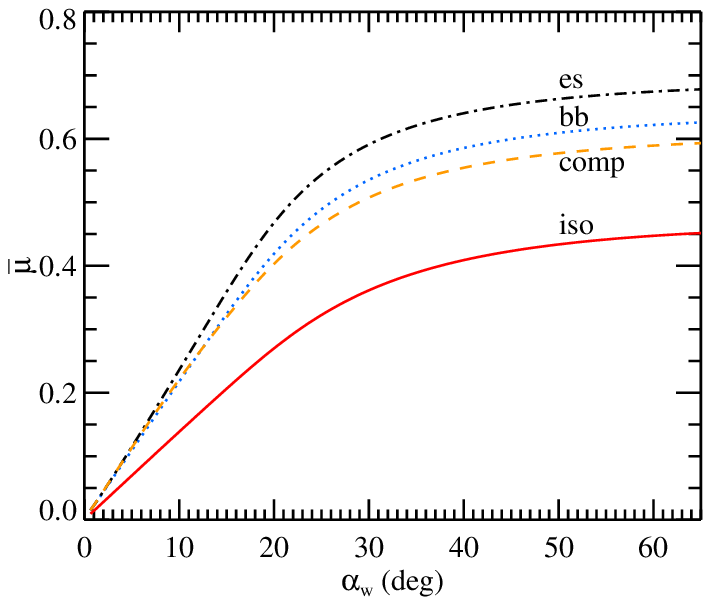}
\caption{Mean interaction angle  as a function of wind opening angle for the four emission patterns:  isotropic (red solid line), Comptonization (orange dashed), blackbody disk (blue dotted), and electron-scattering dominated disk (black dot-dashed). }
\label{fig:muavg}
\end{figure}

\subsection{Properties of scattering medium}

It is evident that the choice of the optical depth profile is critical. 
As an example, we choose a Gaussian profile that is given by
\begin{equation}\label{eq:taumu}
    \tau(\mu) = \tau_0 {\rm e}^{- \mu^2/(2\mu^2_{\rm w})} , 
\end{equation}
where for geometrically thin winds the parameter $\mu_{\rm w}$ can be interpreted as a sine of the characteristic disk opening angle $\alpha_{\rm w}$ but in general it can be even larger than unity. 
Parameter $\tau_0$ is the optical depth along the disk mid plane. 
Although the profile Eq.~\eqref{eq:taumu} describes the optical depth of the wind  and for small $\tau_0$ is equal to the scattering probability of a photon emitted at a given angle, it does not reflect the angular distribution of the central source radiation. 
The effective optical depth defined as a product of the angular emission pattern and the actual optical depth,
\begin{equation}
    \tau_{\rm eff}(\mu) =  a_{\rm inc}(\mu) \tau(\mu) ,
\end{equation}
for small $\tau_0$ is proportional to the probability density $dp/d\mu$ for photons to be scattered as a function of inclination and it is shown in Fig.~\ref{fig:dpdmu}. 
 
For an isotropic source, the effective optical depth is simply the same Gaussian as $\tau(\mu)$. 
The scattered fraction in this case increases with inclination because there is more scattering material close to the disk.  
For the flat, blackbody-like disk, the dependence of the effective optical depth and scattered fraction on $\mu$ is drastically different. 
The effective optical depth being proportional to $\mu$ (since $a_{\rm inc}(\mu)\propto\mu$) implies that viewing the system edge-on ($\mu = \cos i \approx 0$) would only result in the observation of scattered radiation.  
The peak of the effective optical depth is shifted to a larger value of $\mu \sim \mu_{\rm w}=\sin\alpha_{\rm w}$ as compared to the isotropic case and then falls again at lower inclinations where the wind optical depth  decreases. 
A similar dependence of the effective optical depth is seen for the electron-scattering dominated disk. 
Here, the effective optical depth peaks at a slightly higher $\mu$ because the incident emission is more beamed along the disk normal. 
Radiation pattern produced by Comptonization in a slab falls between the considered cases of isotropic and blackbody emission resulting in a peak of $\tau_{\rm eff}(\mu)$ to appear at intermediate values of~$\mu$. 

\begin{figure}
\centering
\includegraphics[width=0.7\linewidth]{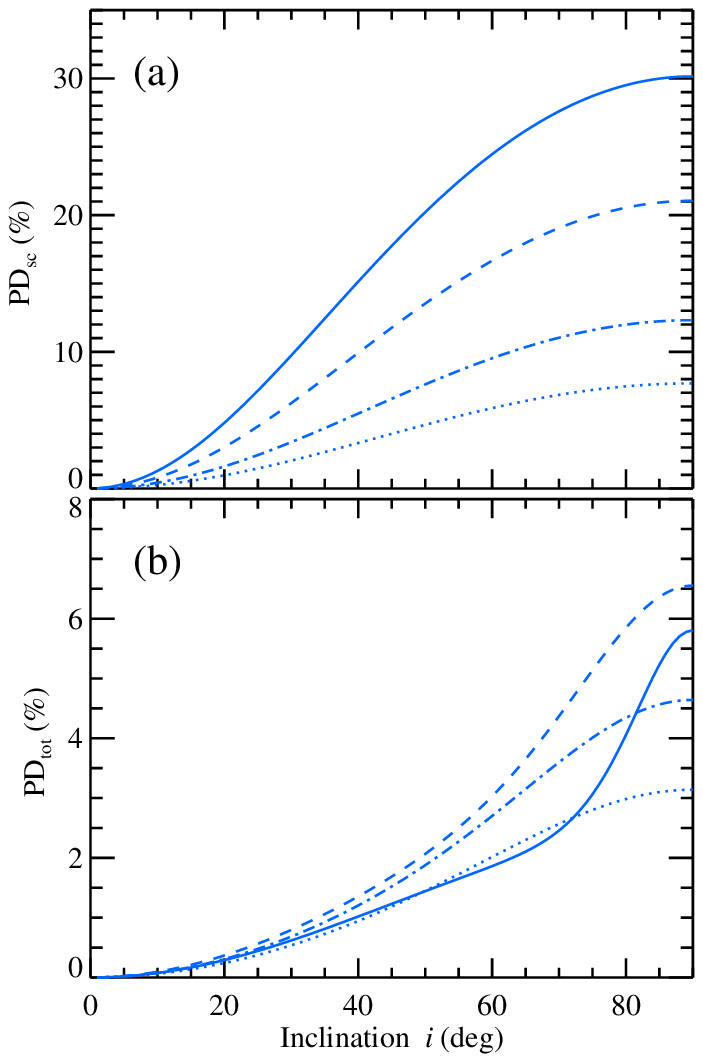}
\caption{PD of (a) scattered component and (b) total emission as a function of inclination for isotropic unpolarized incident source. 
The solid, dashed, dot-dashed, and dotted lines correspond to different opening angles of the wind: $\alpha_{\rm w}=10\degr$, $20\degr$, $30\degr$, and $40\degr$, respectively. 
The equatorial wind optical depth is $\tau_0 = 1$. }
\label{fig:PD_iso}
\end{figure}

The effective optical depth defines the mean cosine of the interaction angle:
\begin{equation} \label{eqn:muavg}
    \Bar{\mu} = \frac{\int_0^1 \mu \tau_{\rm eff}(\mu) d\mu}{\int_0^1 \tau_{\rm eff}(\mu) d\mu} = \int_0^1 \mu \frac{dp(\mu)}{d\mu}  d\mu . 
\end{equation}
It  describes the typical direction of emission of photons which are scattered in the wind for a given type of illuminating source, optical depth profile, and the wind opening angle. 
The variation of $\Bar{\mu}$ with respect to $\alpha_{\rm w}$ is shown in Fig.~\ref{fig:muavg} for our four emission models.  
For the three models, we can also find analytical solutions for the integral in Eq.~(\ref{eqn:muavg}): 
\begin{eqnarray}
\Bar{\mu}_{\rm iso} & = & \frac{1}{\beta} \frac{I_1}{I_0} , \\ 
\Bar{\mu}_{\rm bb} & = &  \frac{1}{\beta} \frac{I_2}{I_1} ,  \\ 
\Bar{\mu}_{\rm es} & = & \frac{1}{\beta} \frac{I_2+I_3\, b/\beta}{I_1+I_2 \, b/\beta } , 
\end{eqnarray}
where $\beta=1/(\mu_{\rm w}\sqrt{2})$, 
$b=2.06$, and integrals $I_n= \int_0^\beta x^n  \exp(-x^2)dx$ are 
\begin{eqnarray}
I_0 & = &  \frac{\sqrt{\pi}}{2} {\rm erf}(\beta), \\ 
I_1 &=& \frac{1}{2} \left[1-\exp(-\beta^2)\right], \\
I_2 &=& \frac{1}{2} \left[-\beta\exp(-\beta^2)+I_0\right], \\ 
I_3 &=& - \frac{\beta^2}{2} \exp(-\beta^2) + I_1,
\end{eqnarray}
and ${\rm erf}$ is the error function. 
We see that all $\Bar{\mu}$ initially increase linearly with $\alpha_{\rm w}$ because of appearance of scattering material at higher latitudes, $\Bar{\mu}_{\rm iso}\sim \sqrt{2/\pi}\sin\alpha_{\rm w}$ and others behave as $\Bar{\mu}\sim \sqrt{\pi/2}\sin\alpha_{\rm w}$. 
However, for  $\alpha_{\rm w}>30\degr$ they saturate at constant values because the wind material becomes more isotropic and the typical interaction angle is defined mostly by the emission pattern. 
For example, for the isotropic wind (i.e. for $\mu_{\rm w}\rightarrow\infty$ and  $\beta\rightarrow0$),   $\Bar{\mu}_{\rm iso}\rightarrow1/2$, $\Bar{\mu}_{\rm bb}\rightarrow2/3$, and $\Bar{\mu}_{\rm es}\rightarrow(1/3+b/4)/(1/2+b/3)\approx0.71$. 
We see that $\Bar{\mu}$ is largest for radiation beamed perpendicularly to the disk (electron scattering-dominated disk) and is smallest for an isotropic source which produces more radiation along the disk where the optical depth of the wind is largest.

\begin{figure}
\centering
\includegraphics[width=0.7\linewidth]{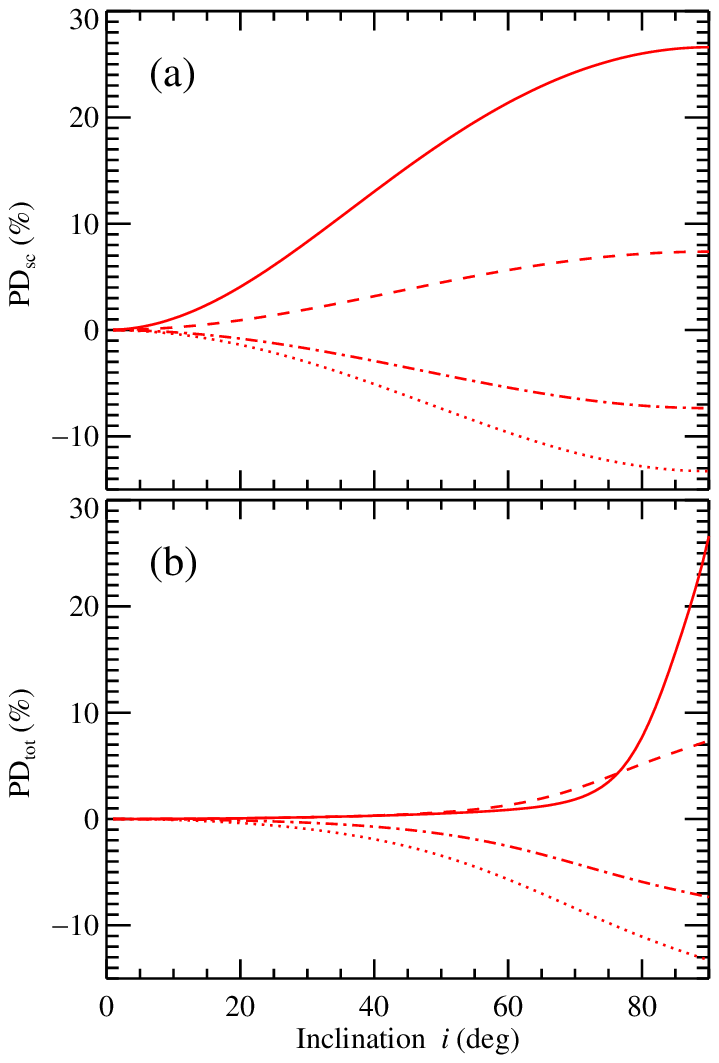}
\caption{Same as Fig.~\ref{fig:PD_iso} but for the blackbody unpolarized disk.}
\label{fig:PD_bb}
\end{figure}

\section{Results}\label{sec:results}

\subsection{Dependence on wind opening angle}
\label{sec:alpha}

Here we study the polarization properties of radiation scattered by the wind depending on the parameters for the four emission models. 
Let us first consider isotropic unpolarized source of incident radiation. 
The dependence of PD of the scattered component and total emission on the inclination of the observer for different opening angles of the wind and fixed $\tau_0=1$ are shown in Fig.~\ref{fig:PD_iso}. 
The PD reduces at lower inclinations because of the higher perceived symmetry of the system. The PD of scattered emission at low $\alpha_{\rm w}$ and high $i$ reaches values of 30\%, in accordance with single scattering in a plane-parallel optically thin slab. 
Substituting $L_*(\mu')=L_*\delta(\mu')$ to Eq.~\eqref{eq:phasematrix}, we get (see Eq.~(18) of \citealt{ST85})
\begin{eqnarray}
\label{eq:Psc}
   P_{\rm sc}(\mu) & = &  
 \frac{\tens{P}_{21}(\mu;0)+
 \tens{P}_{22}(\mu;0) P_{\rm inc}  }{\tens{P}_{11}(\mu;0)+  \tens{P}_{12}(\mu;0) P_{\rm inc} } \nonumber\\ 
& =&  \frac{(1-\mu^2)(1+3P_{\rm inc})}{3-\mu^2 + P_{\rm inc} (1-3\mu^2)},  
\end{eqnarray} 
where $P_{\rm inc}$  is polarization of the incident emission at $\mu'=0$. 
For an unpolarized central source with $P_{\rm inc}=0$, the maximum PD of the scattered radiation reaches 33\% for an edge-on observer. 
It decreases with increasing wind opening angle. This is because at low $\alpha_{\rm w}$, incident radiation is scattered in a plane perpendicular to the disk normal producing positive polarization, while for high $\alpha_{\rm w}$ the mean interaction angle $\Bar{\mu}$ grows, scattering occurs also at high latitudes (high $\mu'$) producing negative polarization and reducing the scattered PD.

The total PD depends on the scattered PD and the fraction of scattered radiation.
We see that maximum possible PD can be reached at a certain value of $\alpha_{\rm w} \sim 20\degr$. 
For lower $\alpha_{\rm w}$, there is less material to scatter while for higher $\alpha_{\rm w}$, the scattering medium becomes more isotropic thereby reducing the polarization.

\begin{figure}
\centering
\includegraphics[width=0.7\linewidth]{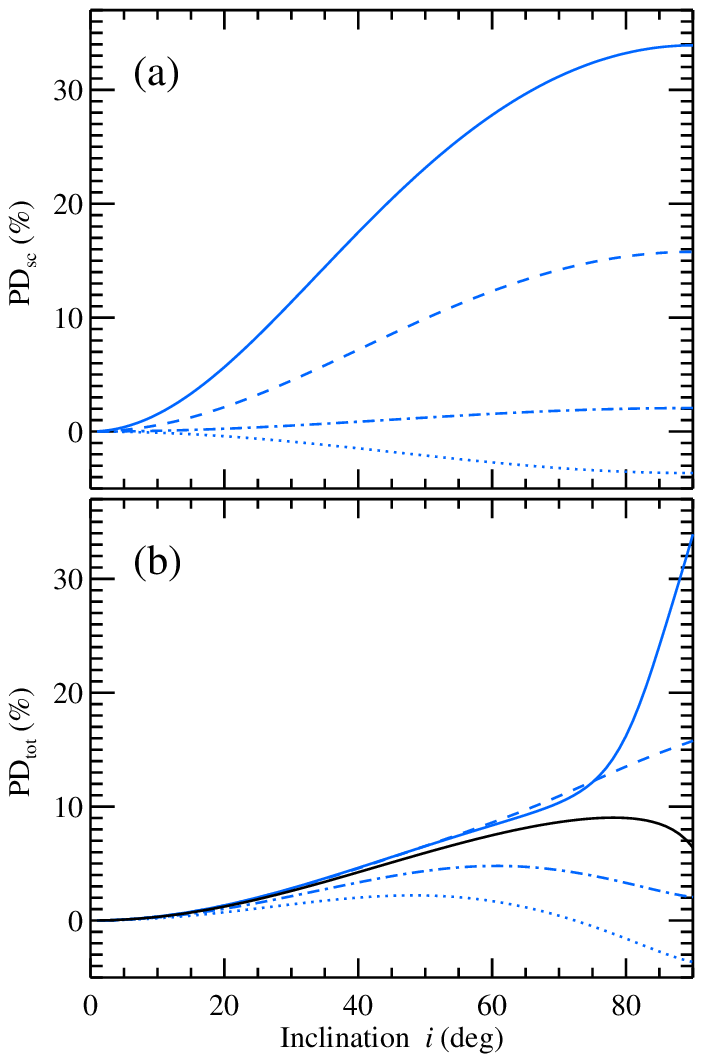}
\caption{Same as Fig.~\ref{fig:PD_iso} but for the Comptonization in a slab. The intrinsic PD of the slab is shown by the solid black line. }
\label{fig:PD_comp}
\end{figure}

The scattered and total PD for the case of the blackbody disk illumination is shown in Fig.~\ref{fig:PD_bb}. Similar to the isotropic case, $\rm PD_{\rm sc}$ is positive for low $\alpha_{\rm w}$ as $\Bar{\mu}$ is small. However, as the angular distribution of the blackbody disk radiation is beamed upwards from the disk plane, the values of $\rm PD_{\rm sc}$ are lower than that of the isotropic case. The vertical beaming of the illuminating radiation and the high $\Bar{\mu}$ values at high $\alpha_{\rm w}$ eventually results in negative (direction perpendicular to disk normal) $\rm PD_{\rm sc}$.  
The total PD is the product of scattered PD and the scattered fraction of radiation. 
The fraction of scattered light is negligible at low inclinations as the source is vertically beamed and almost all of the emission reaching the observer remains unscattered. 
Thus, $\rm PD_{\rm tot}$ is close to zero at low inclinations. 
On the other hand, at very high inclinations, most of the radiation that is reaching the observer is the scattered one. 
The scattered fraction reaches unity for an edge-on observer and $\rm PD_{\rm tot}$ then equals $\rm PD_{\rm sc}$.

\begin{figure}
\centering
\includegraphics[width=0.7\linewidth]{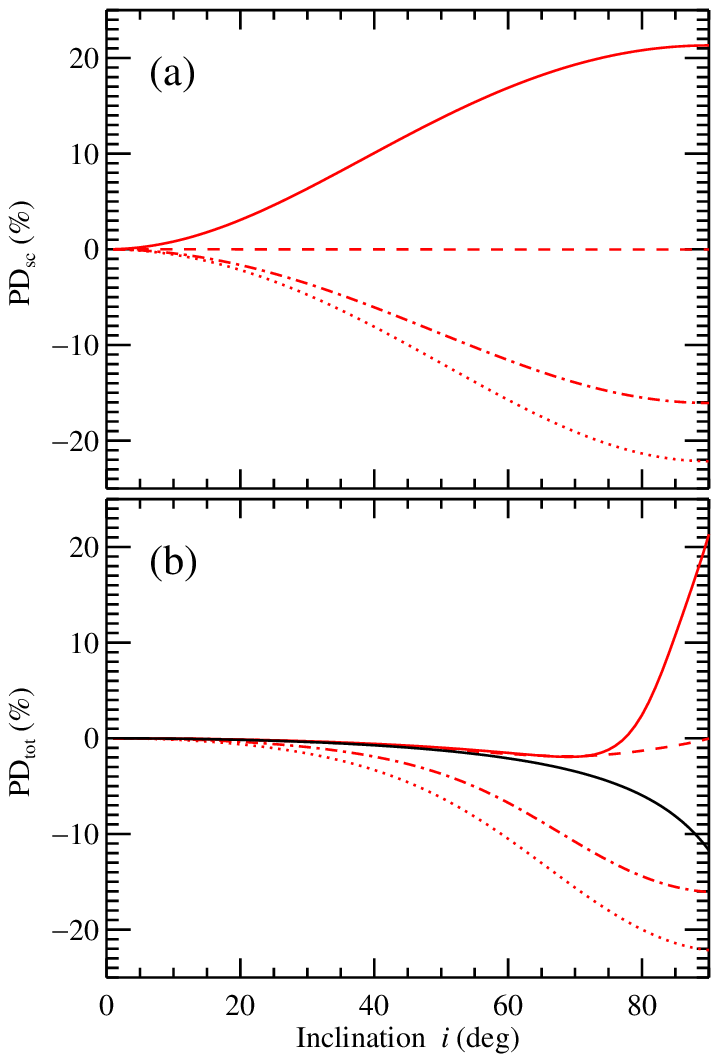}
\caption{Same as Fig.~\ref{fig:PD_iso} but for the electron-scattering dominated disk. The intrinsic PD due to electron scattering in the disk is shown by the solid black line. } 
\label{fig:PD_es}
\end{figure}

In case of a Comptonized slab illuminating source, we see in Fig.~\ref{fig:PD_comp} that for a low $\alpha_{\rm w}$, $\rm PD_{\rm sc}$ is greater than in the isotropic case.
This can be explained by the fact that incident radiation is already polarized in the vertical direction.  
For $P_{\rm inc}=P_{\rm comp}$ of about 7\% (Fig.~\ref{fig:lum_pol}), we see that the numerator in  Eq.~\eqref{eq:Psc} increases by about 20\% comparing to the case of unpolarized incident radiation, while the denominator for $\mu=0$ increases by just $\sim$2\%.
The PD of the total radiation  in this case depends not only on $\rm PD_{\rm sc}$ and the scattered fraction, but also on the intrinsic polarization of the slab. Because the latter is along the same direction as polarization due to scattering in the wind, the two contributions add up. At low inclinations, $\rm PD_{\rm tot}$ is low since both, the scattered and intrinsic PD, are small. $\rm PD_{\rm tot}$ very closely follows the intrinsic polarization of the slab as the fraction of scattered light is very small, thereby reducing the contribution of $\rm PD_{\rm sc}$. At very high inclinations, however, only the scattered component is visible and hence $\rm PD_{\rm tot}$ reaches $\rm PD_{\rm sc}$. 
We see that $\rm PD_{\rm tot}$ exceeds PD of the incident radiation (shown with black solid line) when $\alpha_{\rm w}\lesssim20\degr$. 
For thicker winds, radiation scattered at high latitudes reduces the polarization and even can lead to a sign change. 


The last example is the case of an electron-scattering dominated disk source. The radiation in this case is beamed more strongly in the vertical direction as compared to the blackbody disk case, resulting in a greater $\Bar{\mu}$. 
For small $\alpha_{\rm w}<20\degr$, the PD of the scattered component is positive (Fig.~\ref{fig:PD_es}) while the PD of the incident radiation is negative. 
This results in a reduction of the overall PD. 
On the other hand, for $\alpha_{\rm w}>20\degr$, the PD of the scattered radiation has the same sign as that of the incident one  and for $\alpha_{\rm w}\gtrsim25\degr$ this increases the overall PD.

\begin{figure*}
\centering
\includegraphics[width=0.35\linewidth]{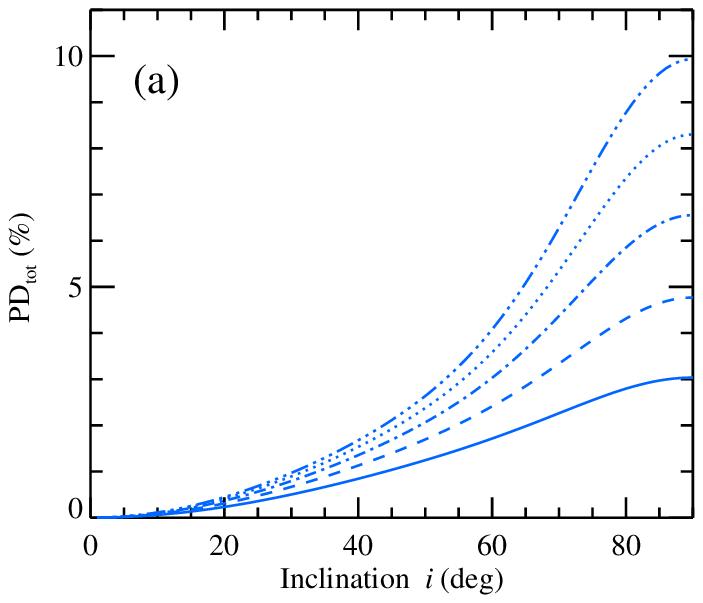}
\hspace{1cm}\includegraphics[width=0.35\linewidth]{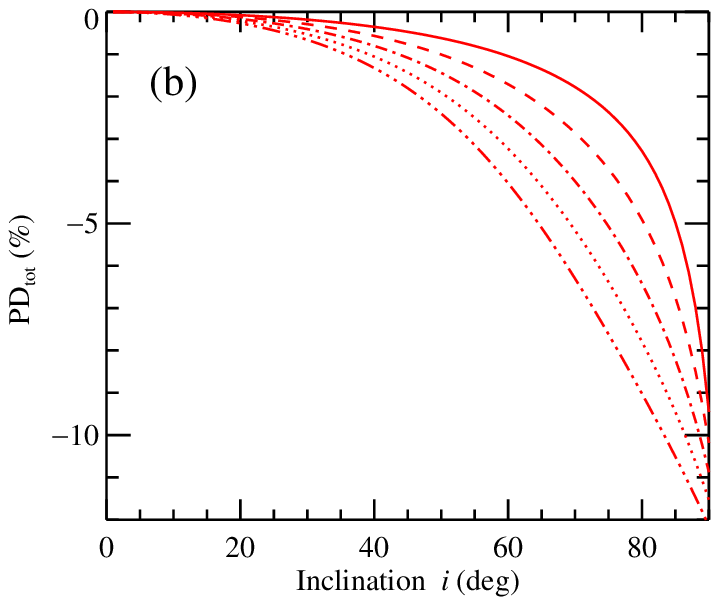}
\includegraphics[width=0.35\linewidth]{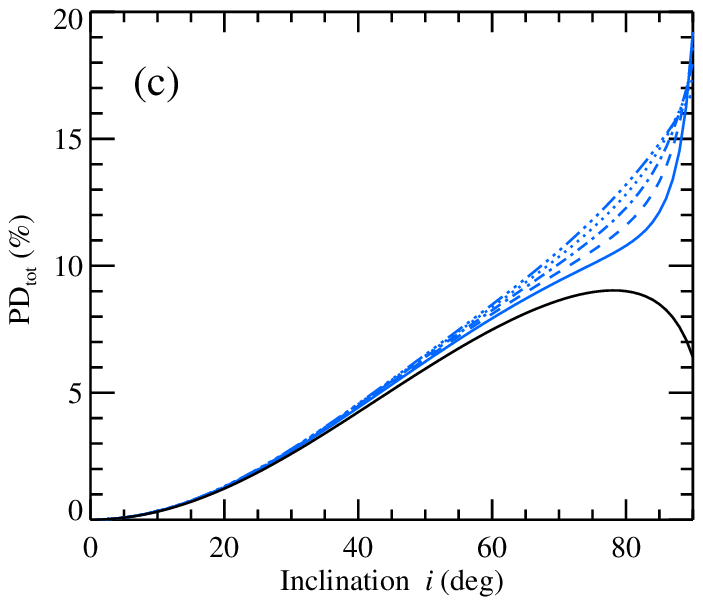}
\hspace{1cm}\includegraphics[width=0.35\linewidth]{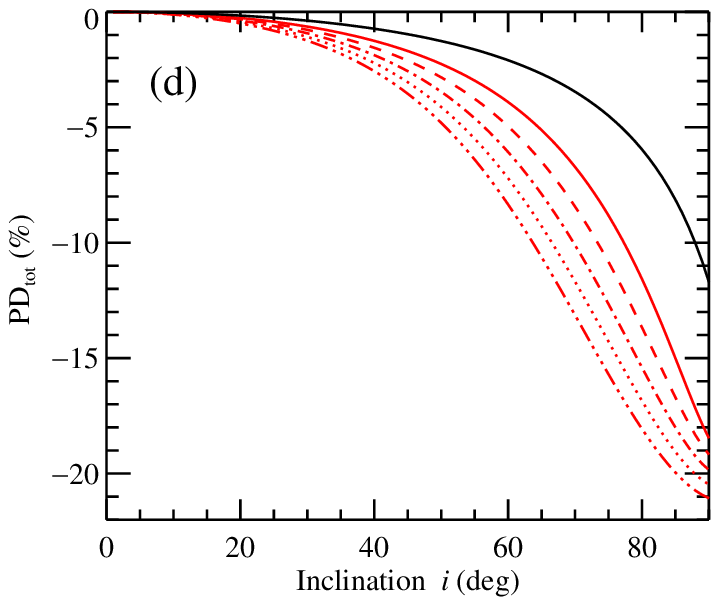}
\caption{Total PD as a function of inclination for varying optical depths: $\tau_0 = $0.5 (solid), 0.75 (dashed), 1 (dotted-dashed), 1.25 (dotted), 1.5 (triple-dot-dashed). The illuminating source is taken to be (a) isotropic, (b) blackbody unpolarized disk, (c) Comptonization in a slab, and (d) electron-scattering dominated disk. 
The wind opening angle $\alpha_{\rm w}$ is fixed at $20\degr$ in cases (a) and (c) and at $40\degr$ in cases (b) and (d). 
In panels (c) and (d), the solid black line shows the PD of the incident  radiation.
}
\label{fig:vary_tau0_all}
\end{figure*}

\subsection{Dependence on wind optical depth}
\label{sec:tau}

\begin{figure*}
\centering
\includegraphics[width=0.4\linewidth]{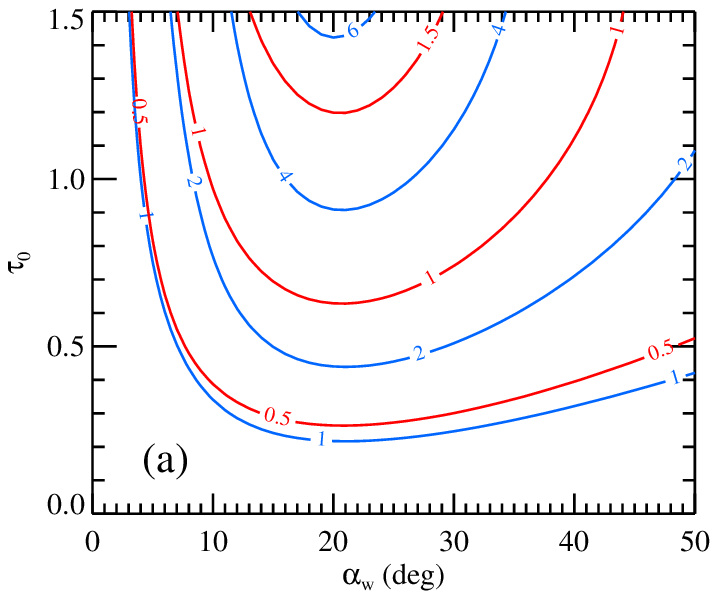}
\includegraphics[width=0.4\linewidth]{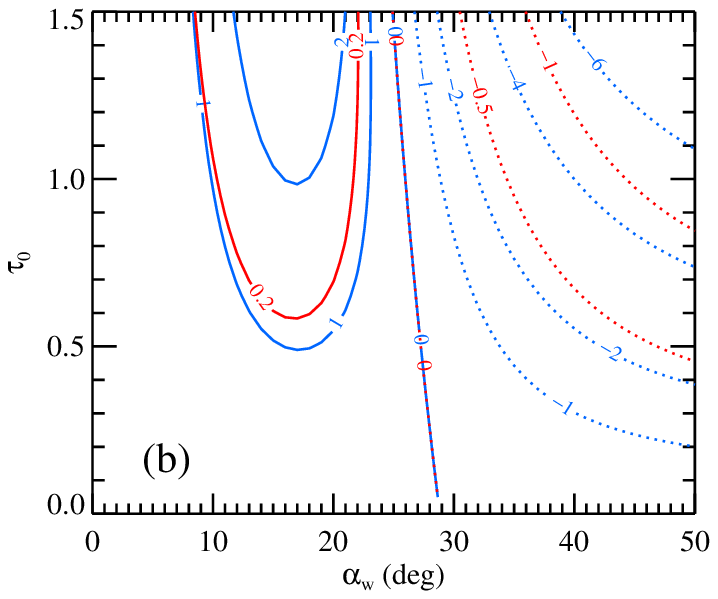}
\includegraphics[width=0.4\linewidth]{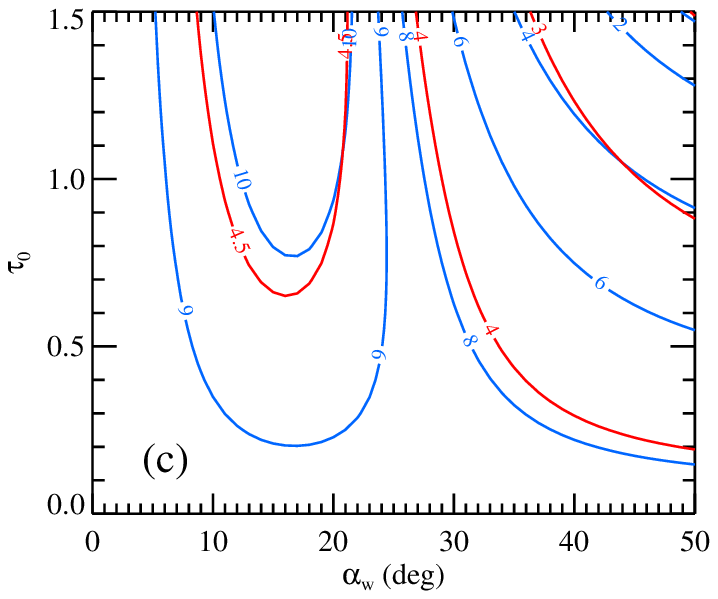}
\includegraphics[width=0.4\linewidth]{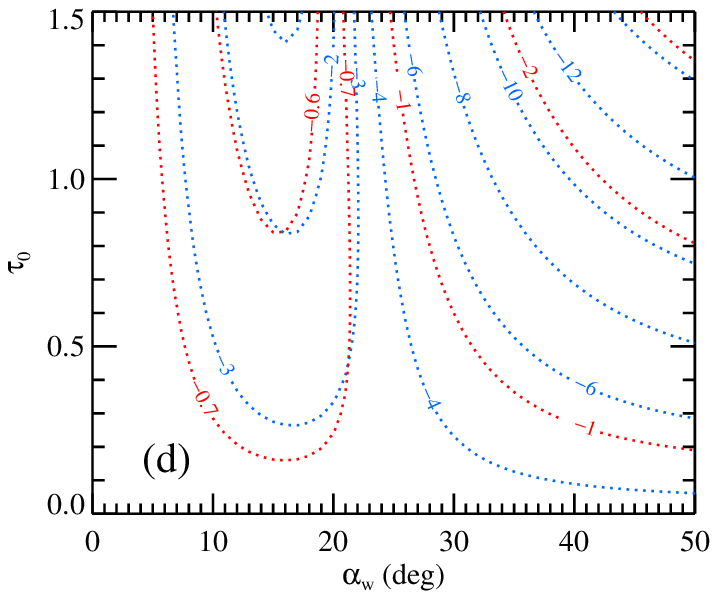}
\caption{Contours of the constant total PD (in \%) on the plane $\tau_0$--$\alpha_{\rm w}$. The red and blue lines correspond to the inclinations of $i=40\degr$ and $70\degr$, respectively. 
The positive PD is shown in the solid lines, while the negative PD with the dotted lines. 
The illuminating source is taken to be (a) isotropic, (b) blackbody unpolarized disk, (c) Comptonization in a slab, and (d) electron-scattering dominated disk. 
The intrinsic PD is zero in cases (a) and (b). 
In case (c), the intrinsic PD is 4.24\% and  8.64\% for the two inclinations 40\degr\ and 70\degr, while in case (d) it is $-0.73\%$ and $-3.46\%$, respectively. 
}
\label{fig:contour_all}
\end{figure*}

The variation of the total PD with inclination for different optical depths for fixed $\alpha_{\rm w}$ is shown in Fig.~\ref{fig:vary_tau0_all}. 
While the highest considered values of $\tau_0$ do not agree with the small optical depth assumption made in our model, it is important to note that $\tau_0$ represents the optical depth value at the disk mid-plane. 
At higher inclinations, the optical depth value is much lower, thus satisfying the assumption.  
The case of an isotropic illuminating source is shown in panel (a). As the value of $\tau_0$ increases from 0.5 to 1.5, $\rm PD_{\rm tot}$ increases because greater optical depth implies larger contribution of the scattered polarized radiation. 
The behavior of $\rm PD_{\rm tot}$ through varying optical depth and wind opening angle is shown in Fig.~\ref{fig:contour_all}(a) for two inclinations $i = 40\degr$ and $70\degr$. We see that the highest PD can be obtained for the highest $\tau_0$ values and for $\alpha_{\rm w} \sim 20\degr$.

For the case of blackbody emission, the optical depth dependence of the PD is shown in Fig.~\ref{fig:vary_tau0_all}(b). 
We see that changing the optical depth does not affect much the maximum attained value of $\rm PD_{\rm tot}$ because we only observe the scattered radiation at high inclinations. 
Varying $\tau_0$, however, leads to changes in the scattering fraction at different inclinations, thereby changing the ${\rm PD}_{\rm tot}(i)$ profile. 
Overall PD on the plane $\tau_0$--$\alpha_{\rm w}$ is shown in Fig.~\ref{fig:contour_all}(b). 
Because the incident radiation here is unpolarized, the sign of the PD is determined by the sign of the PD of scattered radiation, which has  a monotonic behavior with the inclination (Fig.~\ref{fig:PD_bb}).  
We see that for wind opening angles $\alpha_{\rm w} \lesssim 25\degr$, $\rm PD_{\rm tot}$ is positive (vertical), while for larger opening angles, $\rm PD_{\rm tot}$ is negative (horizontal). 
This highlights the importance of the angular distribution of the illuminating source (Fig.~\ref{fig:lum_pol}) and the mean interaction angle (Fig.~\ref{fig:muavg}) to the observed polarization.  

The role of optical depth on the polarization of Comptonizing slab is demonstrated in Fig.~\ref{fig:vary_tau0_all}(c).
We see that the effect of optical depth variation is only important at high inclinations where the contribution of scattered radiation dominates. 
We also see in Figs.~\ref{fig:PD_comp} and \ref{fig:contour_all}(c) that at low inclinations, $\rm PD_{\rm tot}$ has similar values even if $\alpha_{\rm w}$ is varied by a lot, whereas, $\rm PD_{\rm tot}$ varies more strongly with respect to $\alpha_{\rm w}$ at higher inclinations. 
Similarly to the blackbody emission, the PD of the scattered component is positive for $\alpha_{\rm w}\lesssim30\degr$ and negative for thicker winds. 
Because the intrinsic PD is positive in this model, the total PD increases when the PD of the scattered component exceeds the PD of the incident radiation, i.e. for $\alpha_{\rm w}\lesssim25\degr$ and decreases at larger $\alpha_{\rm w}$. 
In particular, the PD of the incident emission  at $i=70\degr$ is 8.64\% and the PD of the total emission may reach values in excess of 10\% for $\tau_0\gtrsim 0.8$ and $\alpha_{\rm w}\sim$ $10\degr$--$20\degr$. 
At $i=40\degr$, the total PD does not reach 5\% independently of the $\tau_0$ and $\alpha_{\rm w}$.  

For the electron-scattering dominated disk, the variations of $\rm PD_{\rm tot}$ due to changes in $\tau_0$ for fixed $\alpha_{\rm w}=40\degr$ is shown in Fig.~\ref{fig:vary_tau0_all}(d). 
We see that similarly to the blackbody disk case, the polarization is negative and  it grows (in absolute value)  with increases $\tau_0$. 
Because the incident radiation here is more beamed perpendicularly to the disk than the blackbody, the typical interaction angle is closer to the disk normal producing higher polarization, which now reaches values $-20\%$ for an edge-on observer. 
The contours of constant  $\rm PD_{\rm tot}$ on the plane $\tau_0$--$\alpha_{\rm w}$ are shown in Fig.~\ref{fig:contour_all}(d). 
For $\alpha_{\rm w}\lesssim20\degr$, the scattered radiation is polarized parallel to the normal, which reduces the total PD. 
On the other hand, for thicker winds  the situation is opposite and  $\rm PD_{\rm tot}$ exceeds PD of the incident radiation. 
The highest polarization is produced by scattering in a geometrically and optically thick wind.

\section{Discussion}
\label{sec:discuss}


\subsection{BH X-ray binaries}
\subsubsection{Hard state}

The low/hard state in BH XRBs is seen via spectra that peak at $\sim$100~keV with a low-temperature disk component \citep{Done07}. The dominant emitting region is associated with a Comptonizing medium \citep{ST85,PS96} of high electron temperatures \citep{Gierlinski1997,IPG05}. 
IXPE observations of the BH binary Cyg X-1 in the hard state measured a PD of $\approx4$\% \citep{Krawczynski22} which is much higher than expected for the inclination of $\lesssim30\degr$ \citep{Miller-Jones2021}. 
This has been explained either with the inner disk being more inclined than the outer disk \citep{Krawczynski22} or through Comptonization in relativistically outflowing corona \citep{Poutanen2023}. 
The PA was found to align with the position angle of the radio jet \citep{Miller-Jones2021}.

Another BH XRB that was observed by IXPE in the hard-intermediate state was Swift J1727.8$-$1613. 
It also showed a PD of $\approx 4$\% and an X-ray PA \citep{Veledina2023} that was aligned with the PA in sub-millimeter and optical wavelengths \citep{Vrtilek2023atel,Kravtsov2023atel} as well as the jet direction \citep{Wood2024}. 
Both of the above mentioned sources show an increasing PD with energy and a constant PA through 2--8 keV range. 
Moreover, the PA in both cases suggest that the hot emission region is likely extended in a plane perpendicular to the disk axis \citep{Krawczynski22,Veledina2023}. 
This disfavors theoretical models  where the coronal geometry is assumed to be spherical or lamppost-like \citep{Dovciak2004,SchnittmanKrolik2010}. 
Instead, models of polarization based on Comptonization of disk or internal synchrotron photons in accretion disk corona or hot flow  \citep{ST85,PS96,SchnittmanKrolik2010,Poutanen2014}, where the hot plasma is extended along the disk plane, are favored.  
However, the high observed PD in these moderate inclination sources still remain unexplained through these models compelling to question whether the inclination of the inner accretion disk is higher than that of the outer accretion disk. 
This can be tested through optical polarization measurements as, for example, in the case of MAXI~J1820+070 \citep{Poutanen2022} where misalignment between the jet axis (BH spin axis) and the orbital axis was discovered. 
Another possibility is that there is a different source of polarization production such as the accretion disk winds. 

The isotropic source (Fig.~\ref{fig:PD_iso}) and the Comptonization  source (Fig.~\ref{fig:PD_comp}) can produce 4\% polarization at inclinations as low as $35\degr$.  
Moreover, Swift J1727.8$-$1613 saw a drop in PD by 1\% as it moved from the hard state to the soft state at high luminosity \citep{Ingram2024} but again recovered its hard state polarization as it traversed back at low luminosity \citep{Podgorny2024}. The complete recovery of polarization properties despite the two order of magnitude difference in luminosity can be described by varying optical depth and the wind opening angle (Fig. \ref{fig:contour_all}).  

Confirmation of polarization production in the winds can be provided by the future \textit{eXTP} mission which offers high temporal resolution X-ray polarimetry \citep{Zhang2019eXTP}. For example, many X-ray transients show QPOs at subsecond frequencies \citep{vanderKlis2005}, and if the X-ray polarization is produced close to the central X-ray source, then this variability should also be reproduced in the polarimetric data. On the other hand, if polarization is produced at much larger distances away from the X-ray source, such as the winds, any variability of the X-ray emission is expected to be smeared out in its polarimetric measurements (see Fig.~4 of \citealt{VeledinaPoutanen2015} for a similar analysis on multi-wavelength QPOs). Thus, one can distinguish between different components contributing to X-ray polarization in different sources.

\subsubsection{Soft state}

The high/soft state in XRBs show a dominant component at $\sim$10~keV and have a non-thermal high energy emission component beyond $\sim$500~keV \citep{ZG04}. The soft component is understood to be thermal emission from an optically thick accretion disk \citep{SS73,NT73}. Both \mbox{Cyg~X-1} and Swift~J1727.8$-$1613 were also observed with IXPE in their soft state \citep{Steiner2024,Svoboda2024swift} where they show a much lower PD ($\sim$1\%; \citealt{Dovciak2024}) than in the hard state. The energy dependence of polarization (that is, the increase in PD with energy and constant PA with energy) as seen in the hard state was maintained in the soft state as well for \mbox{Cyg~X-1} \citep{Steiner2024} and was also seen in the soft state observations of other sources like \mbox{LMC~X-3}  \citep{Svoboda2024lmcx-3} and \mbox{4U~1957+11} \citep{Marra2024}. 
The PD of the latter two sources, even in the highest energy bins, was lower than $\sim$6\% (see Fig.~7 in \citealt{Svoboda2024lmcx-3}). 
In contrast, IXPE observations of \mbox{4U~1630$-$472} in the soft state \citep{Ratheesh2024} showed an average PD of 8.3\% and it rises from  $\sim$6\% at 2~keV to $\sim$10\% at 8~keV.

Pure electron scattering in the disk atmosphere \citep{Cha60,Sob63} can explain the PD seen in moderately inclined sources like \mbox{LMC~X-3}. 
However, being a Thomson scattering process, the PD is constant with energy unlike what is observed. 
Moreover, low inclination sources like \mbox{Cyg~X-1} and \mbox{LMC~X-1} require the incorporation of absorption effects to the accretion disk emission to produce the observed PD. 
Even in this case, the PD tends to decrease with energy (rather than increase as suggested by the observations) in the 2--8 keV band and any spin of the BH can further decrease the PD \citep{Taverna2021}. 
Furthermore, accounting for special and general relativistic effects \citep{StarkConnors1977,Loktev2022,Loktev2024}, one finds the PD to be at most $\sim$6\%, solely for high inclinations. 
For inclinations $< 30\degr$, the PD that is theoretically expected is $<$0.5\%. 
Thus, the general expectations of polarization values fail to explicate the observed PD and its energy dependence in multiple sources observed in the soft state, not including the exceptional case of \mbox{4U~1630$-$472}.

In our models, the blackbody disk source (Fig.~\ref{fig:PD_bb}) and the disk source with electron scattering (Fig.~\ref{fig:PD_es}) characterize the soft state. 
Firstly, we see that sources with inclinations $> 30\degr$ have a PD of about a few percent as observed by IXPE. 
\mbox{Cyg~X-1} seems to have a dominant Comptonized component even in the soft state \citep{G99} and results presented in Fig.~\ref{fig:PD_comp} may characterize this source better. 
Secondly, the PD increases rapidly for higher inclinations and the $\sim$8\% PD seen in 4U~1630$-$472 is realized at inclination $> 60\degr$. 
The estimated inclination of 4U~1630$-$472 is $65\degr < i < 75\degr$ \citep{Kuulkers1998,Tomsick1998}. 
We see that 10\% polarization, perpendicular to the disk normal, can be achieved at $i = 70\degr$ for $\alpha_{\rm w} = 40\degr$. 
Moreover, the X-ray polarization for this source was again measured in its steep power-law (very high) state. 
This usually occurs during a transition between hard to soft states. 
From the $\sim$8.3\% in the soft state, the source PD reduced to $\sim$7.5\% and then $\sim$6.5\% through the steep power-law state \citep{RodriguezCavero2023}. 
This transition can be described by the reduction in wind opening angle, $\alpha_{\rm w}$. 
As $\alpha_{\rm w}$ goes from $40\degr$ to $30\degr$ (corresponding to a decrease in the effective optical depth from 0.63 to 0.55 at $i = 65\degr$), Fig.~\ref{fig:PD_es} shows that for the given inclination, the magnitude of PD decreases from 8\% to 5\%.  
Thus, a gradual decrease in the wind opening angle can automatically describe the change in PD during the state transition period. 
Our models suggest that the polarization observed in 4U~1630$-$472 in its soft and steep power-law state is directed perpendicular to the disk normal.
 
Many of the sources, including 4U~1630$-$472, showed a dependence of PD that increases with increasing energy \citep{Ratheesh2024,Svoboda2024lmcx-3,Steiner2024,Marra2024}. 
While detailed analysis of the energy dependence of PD is left to future work, we discuss here potential origins of such a trend. 
The first could be the influence of local absorption. 
While Thomson scattering is energy independent, true absorption depends on energy. 
Thus, the albedo, $\lambda_E = \sigma_{\rm es}/(\sigma_{\rm es} + \kappa_{\rm ab})$ (where $\sigma_{\rm es}$ is the scattering coefficient and $\kappa_{\rm ab}$ is the true absorption coefficient)  depends on energy and directly influences the scattered fraction. 
Alternatively, inherent energy dependence of the central illuminating source could also lead to energy dependence of resulting polarization. 
\citet{Loktev2022} and \citet{Loktev2024} describe the accretion disk properties including Stokes $I$ and $Q$ (and $U$) parameters for each inclination and energy band after accounting for special and general relativistic effects in the Schwarzschild and the Kerr metrics, respectively. 
Doppler effects due to rotation of the accretion disk can also influence the observed energy dependence of the PD. 
The velocity of the accretion disk increases as the radius decreases resulting in a stronger Doppler boosting: the higher energy photons tend to be beamed more along the disk plane and therefore have a higher chance to be scattered by equatorial wind.  
This effect will lead to a PD growing with energy. 


The absence of energy dependence of PA seen in these above stated sources 
can be explained by scattering in disk winds that are present far away from the compact object thereby being minimally affected by gravitational effects that tend to rotate the PA  \citep{StarkConnors1977,Loktev2022,Loktev2024}.

\subsection{Weakly magnetized NS binaries}

In case of NS XRBs, soft emission can arise from the NS surface or the accretion disk \citep{SS73,SS88}. 
These can be approximately modeled as isotropic or blackbody disk-like emission of the central illuminating source, respectively. 
In principle, the NS surface and the disk can be modeled as a set of blackbodies with different PDs and PAs, and the total polarized flux can be computed from a sum of corresponding Stokes parameters. 
On the other hand, the harder, Comptonized X-rays are emitted either by a hot corona, by the boundary layer (BL) between the NS and the accretion disk, or by the spreading layer (SL) at the NS surface \citep{SS88,IS1999,PS2001,SP2006}. 

IXPE observations of weakly magnetized NS-LMXBs have measured significant polarization from these sources \citep{Ursini2024}. 
PD and PA measurements of sources like GS~1826$-$238 \citep{Capitanio2023} and Cyg~X-2 \citep{Farinelli2023} favor spherically or vertically extended geometry for the optically thick medium rather than one that is radially extended along the disk plane. 
Atolls observed in the soft state -- GX~9+9 \citep{Ursini2023GX9+9}, 4U~1820$-$303 \citep{DiMarco2023}, 4U~1624$-$49 \citep{Saade2024} -- show an increasing trend of PD with energy, similar to what was seen in many BH XRBs. 
While most atoll sources showed a  PD less than $\sim$2\% in the 2--8 keV band, 4U~1820$-$303 and 4U~1624$-$49 showed much higher PD in the higher energy bands (reaching up to $\sim$10\% and $\sim$6\%, respectively; \citealt{DiMarco2023,Saade2024}). 
In case of Z-sources, observations of XTE~J1701$-$462 and GX~5$-$1 have revealed a pattern of PD based on its position on the Z track \citep{Ursini2024}; higher PD of $\sim$4--5\% being seen at the horizontal branch and lower PD of $\sim$1--2\% at the normal and flaring branch \citep{Cocchi2023,Fabiani2024}. 

Expected PD from the disk and the BL are described in, for example, \citet{Dovciak2008} and \citet{Loktev2022}. About $\sim$6--8\% PD is achieved in these models at the highest inclinations, zero spin and at low energies (well below the IXPE range). The contribution to polarization from the SL is modeled by \citet{Farinelli2024} and \citet{Bobrikova2024SL} with a maximum PD of $\sim$2\% at the highest inclinations and high energies. 
Polarization of a few percent seen in the above mentioned sources can be explained by these models. 
However, the high PD seen in 4U~1820$-$303 and 4U~1624$-$49 are still not well understood and  scattering in a wind might need to be invoked. 
In case of 4U~1624$-$49 \citep{Saade2024}, at the source inclination of $>$60\degr\  \citep{frank1987light}, we see that PD of 6\%  can be achieved in both cases: whether the central illuminating X-ray source is taken to be the inner accretion disk (Fig.~\ref{fig:PD_es}) or isotropic (mimicking the SL, for example; see Fig.~\ref{fig:vary_tau0_all}(a)). The direction of the PA with respect to the disk normal will discern between the two possibilities. 
4U~1820$-$303 is a moderate inclination source and the polarization of the hard and soft components are orthogonal to each other \citep{DiMarco2023}. 
The latter property is readily explained by the choice of illuminating source in our wind scattering models; cooler radiation from a disk-type source (Figs.~\ref{fig:PD_bb} and \ref{fig:PD_es}) will have polarization perpendicular to the disk normal, while the hotter radiation from the SL (isotropic) or inner hot flow (Comptonized)  (Figs.~\ref{fig:PD_iso} and \ref{fig:PD_comp}) will be polarized along the disk normal. 
So far in our analysis, we have only considered contribution to the polarization from scattering in the winds above the accretion disk midplane. 
However, additional contribution can also be provided by the winds on the opposite side of the disk, i.e. below the disk midplane (see Fig.~\ref{fig:geometry}). 
Because polarization from both parts will be aligned, greater PD values can be obtained explaining high observed polarization values in sources like 4U~1820$-$303.

Apart from the atoll- and Z-sources mentioned above, IXPE measured the polarization of some peculiar sources like \mbox{Cir~X-1} \citep{Rankin2024} and \mbox{GX~13+1} \citep{Bobrikova2024,Bobrikova2024new} as well. While the PD of \mbox{Cir~X-1} was $\sim$1\% throughout, the PA seemed to vary across different orbital phases ($\sim 50\degr$ rotation) and different hardness ratios \citep{Rankin2024}. 
This has been explained by the presence of two components  responsible for the observed polarization -- BL and SL -- depending on the accretion rate. A misalignment between the NS rotation axis and the orbital disk axis could explain the varying PA. 
Unless the source is highly inclined, the SL cannot produce a polarization of $\sim$1.5\% \citep{Bobrikova2024SL} but if the emission from the SL is scattered in the accretion disk wind, the observed PD can be reached for inclinations as low as $25\degr$ (Fig.~\ref{fig:vary_tau0_all}(a)). 
The absence of strong gravitational effects in wind scattering can explain the difference in PA with respect to polarization produced in the BL. 

GX~13+1 is an X-ray burster that shows variable polarization with respect to both, time and energy \citep{Bobrikova2024}. The PA slowly increased with time, varying up to 70\degr, while the PD first decreases and then increases after an observed dip in the light curve. A constant polarization component, such as scattering in the wind, along with a variable component that depends on the BL and SL were used to explain the observed behavior of polarization.  
The constant component needs to have a PD of $\sim$2\% for a source with high inclination, 60\degr--85\degr (see Fig.~12 in \citealt{Bobrikova2024}). This is easily reached in our models and the direction of PA will help further identify the nature of the central illuminating X-ray source in the case. On the other hand, if we assume that the BL/SL contribute to the constant component, and the winds dictate the variable component, we will need to explain a PD variation from $\sim$5\% to $\sim$3.5\%. This can be understood by the varying optical depth of the wind material. Fig.~\ref{fig:vary_tau0_all}(a) shows that by varying the optical depth, $\tau_0$, and hence the optical depth profile over all inclinations, the total PD produced can be altered. Fixing the inclination at 60\degr, for example, we can obtain 3\% polarization for $\tau_0 = 0.75$ and 5\% for $\tau_0 = 1.25$. 

Another crucial point is the changing PA with time. This can be interpreted by invoking the misaligned BL and the changing optical depth in the winds. Optically thin winds will allow more polarized light from the BL/SL to pass through while optically thicker winds will reduce the light from the central source. Hence the PA of the observed radiation -- that depends on the source of polarization and its geometry -- varies. Further observation and analysis of this source by \citet{Bobrikova2024new} suggests a disk component with PD of $\sim$6.5\% (see their Fig.~5). This is difficult to reach in the IXPE range \citep{Loktev2022} but scattering in the winds provide an alternate solution to reach such high PD values. 

\subsection{X-ray pulsars}

During the first two and a half years of operation, IXPE has observed a dozen of X-ray pulsars \citep[see][for a review]{Poutanen2024Galax}. 
In most of them  polarization was significantly detected and the variations of the PA with the pulse phase followed the rotating vector model \citep[RVM;][]{Meszaros88,poutanen20}.  
However, two bright transients \mbox{LS~V~+44~17} \citep{Doroshenko2023} and \mbox{Swift J0243.6+6124} \citep{Poutanen2024} showed peculiar behavior. 
Two observations of \mbox{LS V+44~17} in 2023 about two weeks apart  showed dramatically different dependence of the PA with pulse phase, with the PA profiles in the two observations being in anti-phase. 
The data could be explained by a two-component model in which one, pulse-phase-dependent component produced by the pulsar follows the RVM, while another polarized component does not vary with phase. 
The polarized flux of the constant component was estimated to be about 3--4\% of the average flux, which, for example, could be produced if that component has a PD of 30\% and contributes 1/10 of the mean flux. 
The suggestion put forward by \citet{Doroshenko2023} was that scattering in the equatorial disk wind could be the origin of the constant polarized component. 
For a high inclination source such as \mbox{LS V+44~17}, our wind scattering model (see Figs.~\ref{fig:PD_iso} and \ref{fig:vary_tau0_all}(a)) can easily reproduce the data.
The value of PD and direction of PA of the constant component not only sheds light on the relevant central X-ray source, but also  constrains the inclination of the source. 
We see that the disk inclination needs to be $\gtrsim 50\degr$ to give a PD of 4\% in the total emission (Fig. \ref{fig:vary_tau0_all}(a)). 

A similar analysis was also done for \mbox{Swift J0243.6+6124} \citep{Poutanen2024}. 
IXPE observed the source three times during its 2023 outburst. 
In two observations,  the pulse phase dependence of PA  followed a double sine wave which is inconsistent with the RVM. 
Using the two-component model, the authors showed that the pulse-phase variation of the PA can be reconciled with the RVM. 
In this source, the polarized flux of the constant  component varied between 1.5\% and 3\% of the mean flux, which  can also be easily explained by the wind scattering. 

We again recall that our models only present polarization produced in winds that are on the side of the disk that is towards the observer (above the disk midplane in Fig.~\ref{fig:geometry}). Contribution from winds on the opposite side of the disk (below the disk midplane) would increase the observed PD. 
Because radiation from X-ray pulsars depends on the azimuthal angle (i.e. on pulsar phase), polarization produced by wind scattering should also vary with pulsar phase, however, with the amplitude smaller than the pulsar radiation itself.   
This would be an interesting problem to be solved in the future. 



\subsection{Seyfert 1 galaxies}

IXPE has measured the X-ray polarization of three Seyfert~1 galaxies, namely, IC~4329A \citep{Ingram2023}, NGC~4151 \citep{Gianolli2023,Gianolli2024}, and MCG-5-23-16 \citep{Marinucci2022,Tagliacozzo2023}. The latter shows a marginally significant polarization of $1.6\pm0.7\%$  (corresponding to the upper limit of 3.2\% at the 99\% confidence level). 
At a possible inclination of 30\degr, such a small PD can be produced by thermal Comptonization in a slab geometry, or it can be produced by Thomson scattering of this radiation in a wind of a wide range of wind opening angles and optical depths (Fig.~\ref{fig:PD_comp}). 
Alternatively, scattering in a wind of $\alpha_{\rm w}\sim40\degr$ of intrinsically unpolarized isotropic source can produce such PD (Fig.~\ref{fig:PD_iso}).

The measured polarization of IC~4329A was $3.3\pm1.1$\% and was aligned with the jet \citep{Ingram2023}. 
From our models (Fig.~\ref{fig:PD_comp}), this implies that the inclination of the source is between $\sim$ 20\degr--35\degr, consistent with the inclination expectation from a Seyfert~1.2 galaxy. 

NGC 4151 shows an X-ray PD of $4.5\pm0.9$\%. 
The measured PA=$81\degr\pm6\degr$ in the 2--8 keV band aligns well with the extended radio emission which  is presumably perpendicular to the accretion disk plane. 
We see that scattering in the winds can reproduce this polarization for inclinations over $40\degr$ (Fig.~\ref{fig:PD_comp}) which is the expected  inclination of the central part of a Seyfert 1.5 galaxy. 
\citet{Bentz2022} modeled the emission line variations constraining the broad-line region opening angle of $\approx$57\degr\ (i.e. 33\degr\ if measured from the disk plane) and the observer inclination of 58\degr.  
The observed X-ray PD can then be reproduced by Comptonization in a slab geometry (see dashed lines in Fig.~\ref{fig:lum_pol}), or by scattering of this radiation in a wedge (wind/broad-line region) of opening angle 30\degr\ (Fig.~\ref{fig:PD_comp}), or even by scattering of isotropic unpolarized source in a wind (Fig.~\ref{fig:vary_tau0_all}(a)). 

\section{Summary}
\label{sec:summary}

Models of XRBs and AGN are well described by a multi-temperature black body disk and a Comptonizing corona. However, recent IXPE observations of these objects have not only shown X-ray PD larger than previously understood but also have discovered variations in polarization that have invoked advanced models. Additionally, XRBs and AGN have shown an ubiquitous presence of accretion disk winds, leading to the speculation that scattering in the winds might contribute to the observed polarization in these sources. In this work, we have developed a model for X-ray polarization produced in XRBs and AGN due to single Thomson scattering in the accretion disk winds with variable wind opening angle and optical depth. We show that our model can explain the high levels of polarization observed in these sources. In case of 4U~1630$-$472, a high PD of $\sim$8\% was observed in the soft state which can be achieved by our models for high inclinations. Moreover, the source showed a constant PA with respect to energy which is expected since scattering in the winds take place much farther away from the central BH and general relativistic rotation of PA is negligible. In the case of GX 13+1, a slow rotation of PA of up to $70\degr$ was seen. This can be explained by scattering in winds that are produced in the outer regions of the accretion disk that is misaligned with the BL/SL of the NS. Slowly varying the optical depth of the winds will naturally result in slow rotation of observed PA. Lastly, in the case of \mbox{LS~V~+44~17}, standard RVM model could not explained the dramatic change in observed PA over a short period of time. Instead, assuming that two components contribute to polarization -- a pulse-phase-dependent component that follows the RVM model and a constant component described by scattering in the winds -- the observed data can be explained. Thus, we show that scattering in accretion disk winds are important in understanding the polarization properties, and hence the geometry of the accretion regimes in XRBs and AGN.  More detailed modeling of a source that is not axially symmetric and the study of the energy dependence of polarization in our models is left for future work. 


\begin{acknowledgements}
This research has been supported by the Academy of Finland grant 355672 (AV) and the Ministry of Science and Higher Education grant 075-15-2024-647. 
Nordita is supported in part by NordForsk.
\end{acknowledgements}

%
%
\bibliographystyle{yahapj}
\bibliography{references.bib} 

\end{document}